\documentclass[aps,pre,showpacs,twocolumn]{revtex4}
\usepackage{graphicx}
\usepackage{psfrag}
\usepackage{amsmath}

\newcommand{\bi}{\begin{itemize}}
\newcommand{\ei}{\end{itemize}}
\newcommand{\ba}{\begin{array}}
\newcommand{\ea}{\end{array}}
\newcommand{\bc}{\begin{center}}
\newcommand{\ec}{\end{center}}
\newcommand{\bee}{\begin{eqnarray}}
\newcommand{\eee}{\end{eqnarray}}
\newcommand{\br}{\mbox{\boldmath $r$}}
\newcommand{\oOmega}{\mbox{\boldmath $\Omega$}}

\newcommand{\bU}{\mbox{\boldmath $U$}}
\newcommand{\m}{\mbox{\boldmath $\mu$}}
\newcommand{\z}{\mbox{\boldmath $\zeta$}}
\newcommand{\bF}{\mbox{\boldmath $F$}}
\newcommand{\bT}{\mbox{\boldmath $T$}}
\newcommand{\nnabla}{\mbox{\boldmath $\nabla$}}

\begin{document}

\title{Hydrodynamic orienting of asymmetric microobjects under gravity}
\author{Maria L. Ekiel-Je\.zewska}
\email{mekiel@ippt.gov.pl}
\affiliation{Institute of Fundamental Technological Research,
    Polish Academy of Sciences, \'Swi\c etokrzyska 21, 00-049 Warsaw, Poland}
\author{Eligiusz Wajnryb}
\affiliation{Institute of Fundamental Technological Research,
    Polish Academy of Sciences, \'Swi\c etokrzyska 21, 00-049 Warsaw, Poland}

\date{\today}

\begin{abstract}
It is shown that nonsymmetric microobjects orient 
while settling under gravity in a viscous fluid. 
To analyze this process, a simple shape is chosen: 
a non-deformable `chain'. The chain consists of two 
straight arms, made of touching solid spheres. In 
the absence of external torques, the spheres are 
free to spin along the arms. The motion of the chain 
is evaluated by solving the Stokes equations with 
the use of the multipole method. It is demonstrated 
that the spinning beads speed up sedimentation by 
a small amount, and increase the orientation rate 
significantly in comparison to the corresponding 
rigid chain. It is shown that chains orient towards 
the V-shaped stable stationary configuration. In 
contrast, rods and star-shaped microobjects do not 
rotate. The hydrodynamic orienting is relevant for 
efficient swimming of non-symmetric microobjects, 
and for sedimenting suspensions.
\end{abstract}

\maketitle

\section{Introduction}
In many biological, medical and industrial applications, it is of interest to predict theoretically what is the sedimentation velocity of small conglomerates of micro-particles under gravity in a fluid~\cite{kasper,LW,geller,mondy,CH}, and how the settling speed can be enhanced or decreased, by a suitable modification of the configuration, or directly by a change of the relative motion of the particles. This issue is especially important for mechanisms of effective swimming, recently intensively investigated for biological systems as well as for artificial micro- and nano-swimmers.

Swimming patterns of various microorganisms have been extensively investigated 
experimentally and theoretically \cite{Berg,Goldstein,Taylor,Purcell,Childress,PedleyKessler,Stone,Bibette,Hosoi}. 
Microorganisms propel themselves owing to a periodic change of their shape and possibly also its orientation in space. Often the core cell does not deform, and the shape is changed owing to waving, ondulating or rotating flagella.
Typical sizes of bacteria, spermatozoa 
or algae lie in the range from 1 to 200 $\mu$m, and their swimming speeds are usually 
up to several hundred $\mu$m/s. For such microobjects moving in 
aqueous environments, the fluid inertia and the Brownian motion are irrelevant~\cite{PedleyKessler}. 
Therefore, a theoretical model of swimming should be based on hydrodynamic 
interactions~\cite{kim} 
between individual parts of the microobject, following from the stationary Stokes equations and the appropriate boundary conditions at the surface of the swimmer. Typically, in the swimming problem a periodic sequence of flagella shapes and the corresponding translational and rotational velocities of their parts relative to the core cell are known as functions of time, as well as the total force and torque exerted on the cell center. The task is to determine the translational velocity of the center, and also the cell's angular velocity. There exist a number of models of freely moving swimmers (the total force and torque vanish).

However, the microorganisms are often denser than the water in which they swim, by a few percent for the algae, approximately 10\% for bacteria and 30\% for spermatozoa,
and the mass distribution can be non-uniform~\cite{PedleyKessler}. The gravitational force 
is essential for explanation of orientational mechanisms (gravitaxis) observed experimentally e.g. for algae~\cite{Kessler,Kessler2}. 
In general, hydrodynamic interactions would tend to orient non-symmetric microobjects 
settling under gravity. This effect, certainly important for swimming, will be investigated in this paper. 

We focus on a very simple model: a non-symmetric `chain' 
of three identical spheres, with two pairs at contact, but not the third one. Owing to the lubrication forces, the shape of the conglomerate is fixed. In Sec.~\ref{model}, the accurate spherical multipole method~\cite{fel,cfhwb} of solving the Stokes equations is introduced, with the lubrication correction for the relative motion of close surfaces~\cite{jcp}, and the HYDROMULTIPOLE numerical code~\cite{jcp}. 
In Sec.~\ref{mot}, 
we use this method to evaluate the translational, rotational and spinning velocities of the microobject, determine how it orients while settling under gravity, and find the stable stationary configuration. In Sec.~\ref{sta}, we study how the settling speed depends on shape, by comparing sedimentation velocities of all the stationary configurations of three spheres~\cite{AM,EW,hocking}. We also investigate how accurate is the point-particle approximation~\cite{free}. In Sec.~\ref{con} we summarize the results obtained for the chain made of three spheres. We also check if chains with two arms made of a larger number of beads  orient hydrodynamically, qualitatively in the same way as the simple three-sphere model.

\section{The model of a moving asymmetric microobject}\label{model}
Consider three identical spheres falling under gravitational forces $\tilde{\bF}_0$ in an infinite fluid of viscosity $\eta$.
The low Reynolds number is assumed for the corresponding fluid flow. 
The fluid velocity 
${\bf v}(\br)$ and pressure 
$p(\br)$ satisfy the stationary Stokes equations~\cite{kim,happel},
\bee
\eta {\bf \nnabla}^2 {\bf v} -{\bf \nnabla} p = {\bf 0}, &\hspace{2cm}& 
{\bf \nnabla} \cdot {\bf v} = 0,\label{incompressible}
\eee
with the stick boundary conditions at the surfaces of the spheres and no fluid flow at infinity.
Therefore the translational $\tilde{\bU}_i$ and rotational $\tilde{\oOmega}_i$ 
velocities of each sphere $i=1,2,3$ 
are linear functions of the force $\tilde{\bF}_0$,
\bee
\tilde{\bU}_i &=& \left[ \sum_{k=1}^3 \tilde{\m}_{ik}^{tt}\right] \cdot  \tilde{\bF}_0,\label{em1} \\
\tilde{\oOmega}_i &=& \left[ \sum_{k=1}^3 \tilde{\m}_{ik}^{rt}\right] \cdot  \tilde{\bF}_0, 
\hspace{1cm} i=1,2,3.\label{em2}
\eee
The $3\times3$  mobility 
matrices $\tilde{\m}_{ik}^{tt}$ and $\tilde{\m}_{ik}^{rt}$ are to be found as 
functions of the relative positions $\tilde{\br}_l-\tilde{\br}_j$ of the 
sphere centers.

In the following, as in Ref.~\cite{EW}, particle positions $\tilde{\br}_i$ will be  
normalized by the sphere diameter $d$, translational velocities $\tilde{\bU}_i$ by the Stokes velocity. 
\bee
U_S=|\tilde{\bF}_0|/(3\pi\eta d), 
\eee
angular velocities $\tilde{\oOmega}_i$ by $2U_S/d$, and  time $\tilde{t}$ by two Stokes times, $2\tau_S$, with
\bee
\tau_S=d/(2U_S). 
\eee
 The corresponding  dimensionless quantities are
\bee
\br_i &=& \tilde{\br}_i/d,\label{n1}\\
\bU_i&=& \tilde{\bU}_i/U_S,\\
t &=& \tilde{t}/(2 \tau_S),\\
\oOmega_i &=& \tilde{\oOmega}_i \, \tau_S,\\
{\m}_{ik}^{tt}&=&  \tilde{\m}_{ik}^{tt} \cdot 3\pi\eta d,\\
{\m}_{ik}^{rt}&=& \tilde{\m}_{ik}^{rt} \cdot 3\pi\eta d^2/2 ,\\
\bF_0&=& \tilde{\bF}_0/|\tilde{\bF}_0|,\label{nend}
\eee
and the dimensionless Eqs.~\eqref{em1}-\eqref{em2} read,
\bee
{\bU}_i  &=& \left[ \sum_{k=1}^3\m_{ik}^{tt}\right] \cdot  \bF_0, \label{emd1}\\
{\oOmega}_i  &=& \left[ \sum_{k=1}^3\m_{ik}^{rt}\right] \cdot  \bF_0, 
\hspace{1cm} i=1,2,3.\label{emd2}
\eee

\begin{figure}[ht]
\resizebox{5.6cm}{!}{\includegraphics{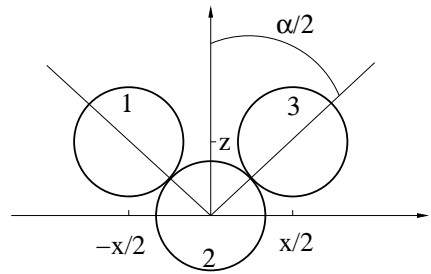}}
\caption{Parametrization of a chain configuration.}\label{skosna}
\end{figure}

In our model it is assumed that each sphere touches another one. Once it happens, the spheres remain at contact owing to lubrication forces~\cite{JO}.
Such configurations will be called `chains'. As illustrated in Fig.~\ref{skosna}, positions of the sphere centers are parametrized by the angle $\alpha$ between the chain links,
\bee
\br_1&=&(-x/2,0,z),\\
\br_2&=&(0,0,0),\\
\br_3&=&(x/2,0,z),
\eee
with $x=2\sin(\alpha/2)$ and $z=\cos(\alpha/2)$. 
Orientation 
of the gravitational force is arbitrary, $\bF_0=(F_{0x},\;F_{0y},\;F_{0z})$.

Evaluation of the mobility matrices of chains requires a
special treatment, because now 
there are at least two contact points. Moreover, it
is essential to specify
more precisely what is meant by ``the contact''. The first possibility 
is to assume that there is no external forces other than gravity 
exerted on each of the spheres. In this case, 
lubrication does not allow for any relative motion of the touching spheres
except their spinning along the line of centers~\cite{JO}. 
In the following, such a contact will be called ``beads''.
The second option is to ``glue'' the touching spheres,
imposing on them external torques, which prevent them from any relative 
motion. Such a contact will be called ``rigid''. In Appendix~\ref{collective}, 
it is explained
in details what is the difference between 
the  mobility matrices for both types of 
the chains.

The essential task of this paper is to determine the 3x3 mobility 
matrices $\m_{ik}^{tt}$ and $\m_{ik}^{rt}$ both for chains made of 
beads and the rigid ones.  Because of hydrodynamic interactions of the close surfaces, we have to go beyond the point-particle approximation, and even beyond superposition of two-body mobilities~\cite{kim,fel}.

Therefore we evaluate the three-sphere mobilities numerically by the multipole expansion  \cite{kim,fel}.  
The algorithm from Ref.~\cite{cfhwb} and its accurate numerical FORTRAN implementation 
HYDROMULTIPOLE~\cite{jcp} are applied.  
The accuracy is controlled by a varied order 
of the multipole truncation $L$ (see Ref.~\cite{cfhwb} for the definition of $L$ 
and Refs.~\cite{cfhwb,accuracy} for discussion of the accuracy estimates).
In this paper, we take a large value  $L=6$. The results will be presented in the next section.

\section{Motion of a chain}\label{mot}
It is convenient to describe the motion of a chain (rigid or made of beads)
referring to its center of mass, ${\bf R}= (\br_1+ \br_2+ \br_3)/3$, because 
in this case the total external torque vanishes. 
The task is to find its translational 
and rotational 
velocities,  
\bee
{\bU} &=&(\bU_1+\bU_2+\bU_3)/3,\\
\oOmega &=& \oOmega_2 =(\oOmega_1+\oOmega_2+\oOmega_3)/3.
\eee
For the chain made of beads, the spinning speed $\omega^{beads}$ also needs to be determined; 
from the symmetry of the system it follows that 
\bee
\omega^{beads} \hat{\br}_{12}&=& \oOmega_1-\oOmega_2, \hspace{0.3cm}\mbox{and} \hspace{0.3cm} -\omega^{beads} \hat{\br}_{32}= \oOmega_3-\oOmega_2,
\nonumber \\
\eee
with the unit vector $\hat{\br}_{ij}=\br_{ij}/|\br_{ij}|$, and the standard notation for the relative positions, $\br_{ij}=\br_i -\br_j$.

\begin{figure}[ht]
\resizebox{5.6cm}{!}{\includegraphics{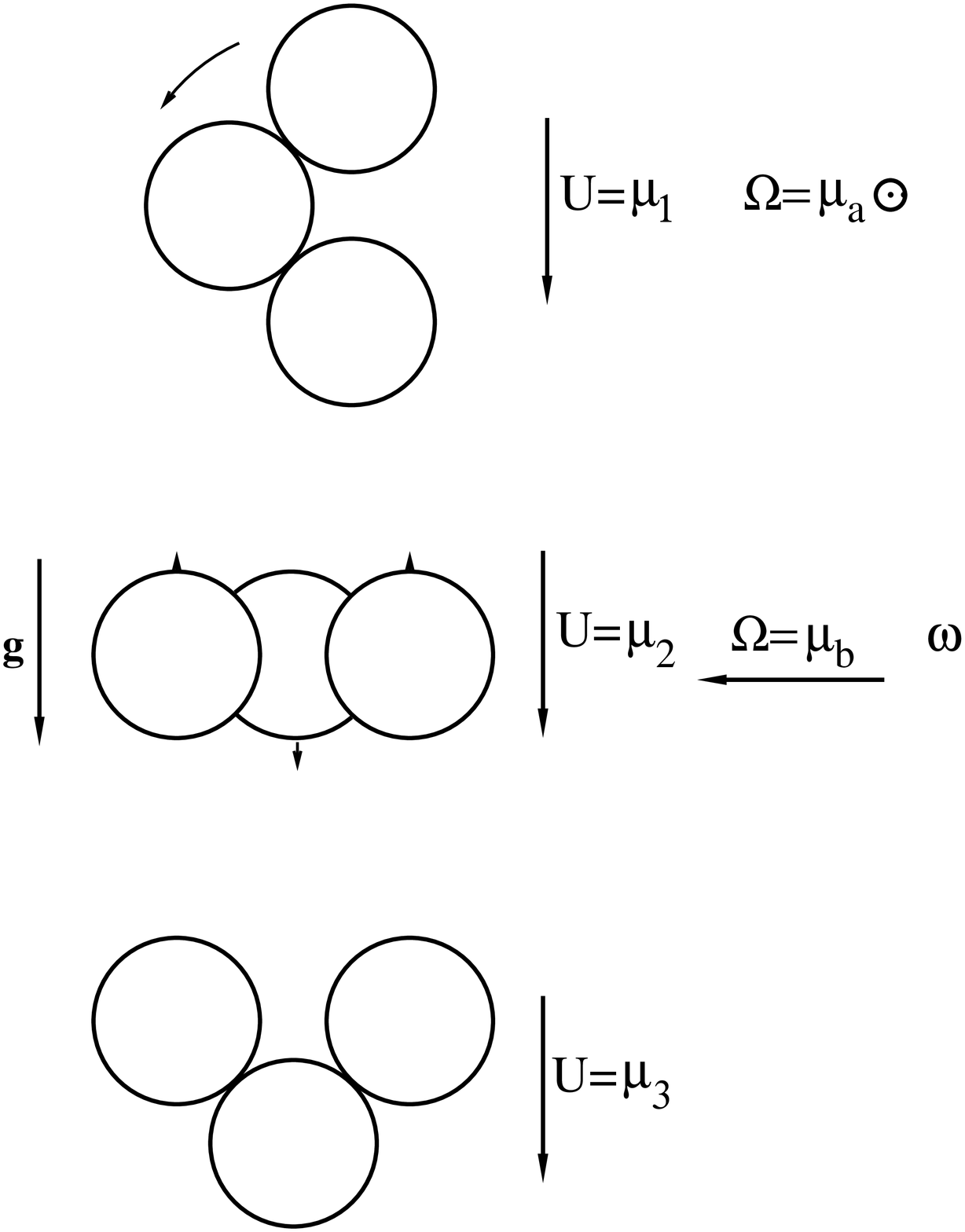}}
 \caption{Degrees of freedom of chains at the characteristic orientations with respect to gravity. Top: gravity along x; middle: gravity along y; down: gravity along z.}\label{in}
\end{figure}

The  translational and angular velocities of the chain depend linearly on the total  gravitational force acting on the chain, $\bF=3\bF_0$, 
\bee
{\bU}&=&\m^{tt} \cdot \bF,\label{uU}\\
\oOmega &=&\m ^{rt}\cdot \bF,\label{oO}\\
\omega &=&\m^{\omega}\cdot \bF, \label{o}
\eee
with the mobility matrices of the chain to be found. Owing to the symmetry, in the frame of reference shown in Fig.~\ref{skosna},
\bee 
\m^{tt} &=& \frac{1}{3} \left(\!
\ba{ccc}
\mu_1(\alpha)&0&0\\
0&\mu_2(\alpha)&0\\
0&0&\mu_3(\alpha)
\ea
\!\!\right),\label{transki}\\
\m^{rt}&=&  \frac{1}{3} 
\left(
\!\!\!\ba{ccc}
0&\!\!\!\mu_b(\alpha)\;\;&\;0\;\\
-\mu_a(\alpha)&\!\!\!0\;\;&\;0\;\\
0&\!\!\!0\;\;&\;0\;
\ea
\right),\label{rotki}\\
\m^{\omega}&=&\;\; \frac{1}{3} \left(\;\;\,
\ba{ccc} 0\;\;\; &\mu_{\omega}(\alpha)\;\; &0\;
\ea
\right).
\eee
In the above equations, the factor 1/3 has been introduced. With this choice,
in the frame of reference shown in Fig.~\ref{skosna}, and with the adopted normalization~\eqref{n1}-\eqref{nend}, 
the  mobility coefficients for the chain are just equal to the corresponding velocity components, 
$\mu_1=U_{x}$, $\mu_2=U_y$, $\mu_3=U_z$, $\mu_a=-\Omega_y$ and $\mu_b=\Omega_x$.
The physical meaning of the coefficients is indicated in Fig.~\ref{in}.

The  mobility coefficients in 
Eqs.~\eqref{transki}-\eqref{rotki} are determined from 
combinations of the three-sphere friction coefficients for the individual spheres, as
 outlined in Appendix~\ref{collective}. In general, the  mobility 
coefficients for a chain of beads differ from those for a rigid chain at the same configuration and in this case they
will be denoted by 
the corresponding superscripts. 
Thus, $\mu_2^{beads}\ne \mu_2^{rigid}$ and $\mu_b^{beads}\ne \mu_b^{rigid}$. 
For a rigid chain, there is no spinning and $\mu_{\omega}^{rigid}=\omega^{rigid}=0$, 
while the chain of beads does spin, $\mu_{\omega}^{beads}=\omega^{beads}\ne 0$
However, 
for a rigid chain and the chain of beads at the same configuration,
$\mu_1$, $\mu_3$ and $\mu_a$ are identical and those coefficients will not be marked by 
any superscripts.

\begin{figure}[ht]
\psfrag{mua}{\huge $\mu_a$}
\psfrag{p}{\huge $\pi$}
\psfrag{p/3}{\huge $\pi/3$}
\psfrag{2p/3}{\huge $2\pi/3$}
\psfrag{m1, m2 and m3}{\huge $\!\!\!\!\!\!\mu_1$, $\mu_2^{rigid}$, $\mu_2^{beads}$ and $\mu_3$}
\psfrag{alpha}{\huge $\hspace{2.6cm}\alpha$}
\psfrag{mu1}{\huge $\mu_1$}
\psfrag{mu3}{\huge $\mu_3$}
\psfrag{mu2r}{\huge $\!\!\mu_2^{rigid}$}
\psfrag{mu2b}{\huge $\mu_2^{beads}$}
\resizebox{8.6cm}{!}{\includegraphics{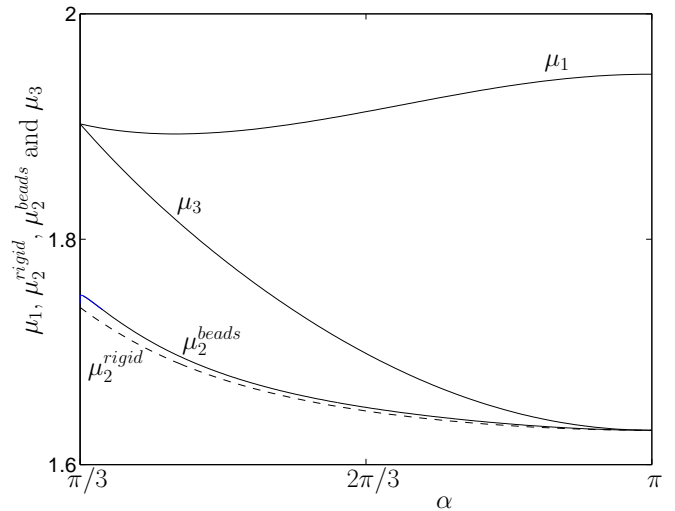}}
\caption{The chain translational velocity components.}
\label{trans}
\end{figure}

The translational motion will now be determined. The corresponding  mobility coefficients  are evaluated numerically and plotted in 
Fig.~\ref{trans} as functions of the angle $\alpha$. For $\alpha=\pi/3$, the sphere centers 
form
the equilateral triangle with three contact points; in the following, this configuration will be called `a star'. For $\alpha=\pi$, the sphere centers are aligned; this configuration will be called `a rod'.
For $\mu_2$ and $\mu_3$, when the base of 
the chain is perpendicular to gravity, then 
its settling velocity is a decreasing function of the apex angle $\alpha$, in agreement 
with the intuitive prediction that stretching the arms should increase the friction force.
For $\mu_1$, when the base of 
the chain is parallel to gravity, it is intuitive to expect the opposite effect: with the 
increase of $\alpha$, the chain aligns along gravity, and its resistance is weaker. This is indeed observed for a wide range of the larger angles, except  those relatively close to $\pi/3$.

For a given shape (fixed $\alpha$), we now compare the magnitude of $\mu_1$, $\mu_2$ and $\mu_3$. Settling is always the fastest if gravity is along the base of the chain, and the slowest if gravity is perpendicular to the plane of the chain.  
Notice that in the last case, the chain made of spinning beads sediments faster than the corresponding rigid one. 

\begin{figure}[ht]
\psfrag{mua}{\huge $\mu_a$}
\psfrag{mubr}{\huge $\mu_b^{rigid}$}
\psfrag{mubb}{\huge $\mu_b^{beads}$}
\psfrag{p}{\huge $\pi$}
\psfrag{p/3}{\huge $\pi/3$}
\psfrag{2p/3}{\huge $2\pi/3$}
\psfrag{m1, m2 and m3}{\huge $\!\!\!\!\!\!\mu_1$, $\mu_2^{rigid}$, $\mu_2^{beads}$ and $\mu_3$}
\psfrag{alpha}{\huge $\hspace{2.6cm}\alpha$}
\psfrag{muw}{\huge $\mu_{\omega}$}
\resizebox{8.6cm}{!}{\includegraphics{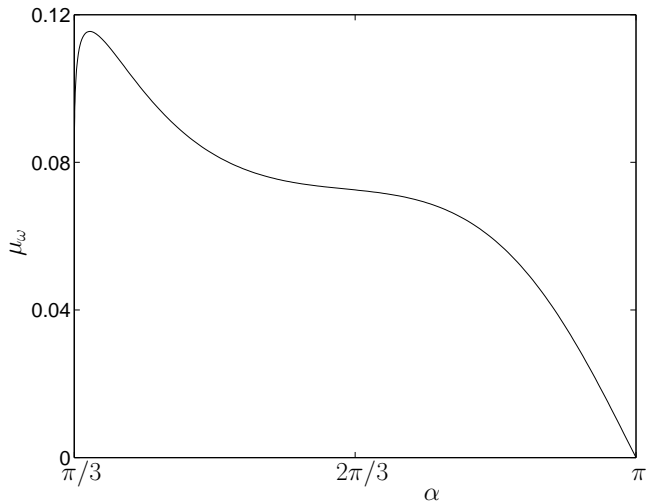}}
\caption{The spinning speed of the chain of beads.}
\label{obrot}
\end{figure}

The spinning speed is plotted in Fig.~\ref{obrot}. It reaches a maximum for a very small value of $\alpha$, and then slowly decreases.
For a wide range of the angles between the chain arms, the spinning surfaces move with velocities which are still around 5\% of the settling speed.

\begin{figure}[ht]
\psfrag{mua}{\huge $\mu_a$}
\psfrag{mubr}{\huge $\mu_b^{rigid}$}
\psfrag{mubb}{\huge $\mu_b^{beads}$}
\psfrag{p}{\huge $\pi$}
\psfrag{p/3}{\huge $\pi/3$}
\psfrag{2p/3}{\huge $2\pi/3$}
\psfrag{mua, mubr and mubb}{\huge $\!\!\!\!\!\!\mu_a$, $\mu_b^{rigid}$ and $\mu_b^{beads}$}
\psfrag{alpha}{\huge $\hspace{2.6cm}\alpha$}
\resizebox{8.6cm}{!}{\includegraphics{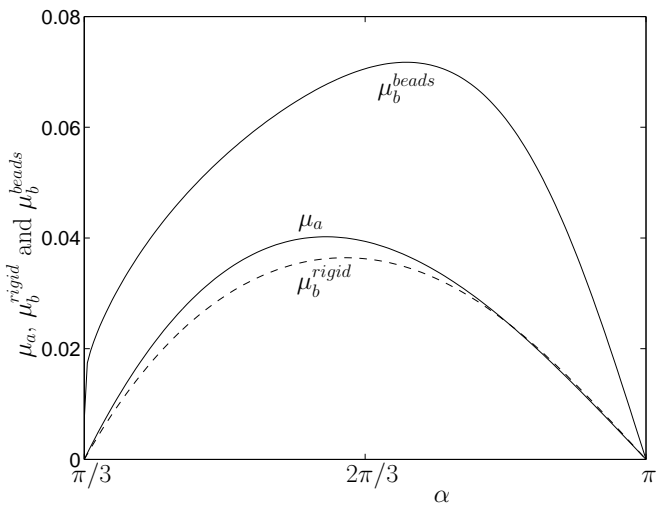}}
\caption{The chain angular velocity components.}
\label{omegi}
\end{figure}

Next, we evaluate the chain rotation. The corresponding 
 mobility coefficients are plotted 
in~Fig.~\ref{omegi} as functions of $\alpha$. 
Each component of the angular velocity is zero for stars and rods, and
has a maximum  at intermediate values of the angle between the arms, smaller than $2\pi/3$ 
for a rigid chain. Therefore the speed of the hydrodynamic orienting is very sensitive to 
shape. It is remarkable that spinning of the beads has a profound effect on the chain rotation. The spinning increases the angular velocity by at least a factor of two in comparison to the corresponding rigid chain at the same configuration. 
Moreover, 
we observe a much wider range of the chain shapes for which the rotation is significant. Indeed,
the maximum is shifted to larger values of $\alpha$, above $2\pi/3$, and there is a qualitative difference at small angles, where the lubrication interactions between the spinning beads keep the chain rotating.

\begin{figure}[ht]
\psfrag{mu/10}{\huge $\mu_{\omega}/10$}
\psfrag{mub}{\huge $\mu_b^{beads}$}
\psfrag{mu1-1.739}{\huge $\!\!\mu_2^{beads}-1.739$}
\psfrag{m1, mb and m}{\huge $\!\!\!\!\!\!\!\!\!\!\!\!\mu_b^{beads}$, $\mu_2^{beads}$ 
and $\mu_{\omega}$}
\psfrag{1/ln(alpha-pi/3)-1}{\huge $-1/\ln (\alpha-\pi/3)$}
\resizebox{8.6cm}{!}{\includegraphics{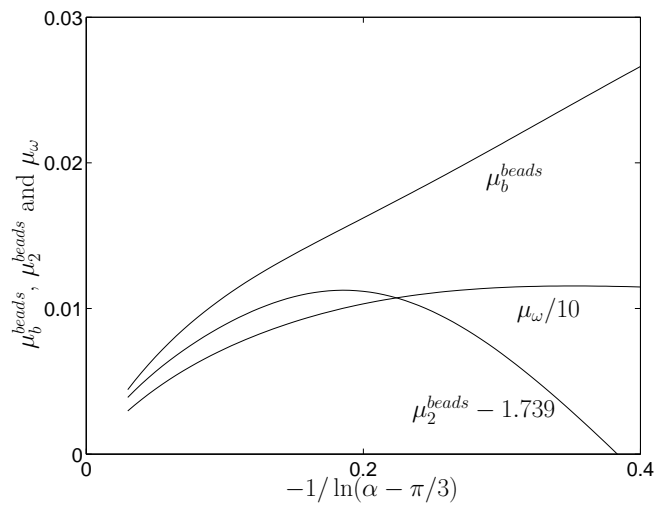}}
\caption{The mobility coefficients for a spinning chain of beads with a small $\alpha$.}
\label{lupa}
\end{figure}

When the close surfaces of spheres 1 and 3 move with respect to each other, then
the mobility coefficients $\mu_2^{beads}$, $\mu_b^{beads}$ and  $\mu_{\omega}^{beads}$  decrease rapidly with the decreasing $\alpha$ only if $\alpha-\pi/3$ becomes extremely small, as seen in  Figs.~\ref{trans},~\ref{omegi}, and~\ref{obrot}. This is the typical lubrication interaction of very close surfaces in relative motion~\cite{JO}. Actually, the relative mobility coefficients decrease to zero as the inverse logarithm of the gap size. This scaling can be seen in Fig.~\ref{lupa}, where $\mu_2^{beads}$, $\mu_b^{beads}$ and $\mu_{\omega}^{beads}$ are plotted 
as functions of $[-1/\ln(\alpha-\pi/3)]$. In fact, to account for the relative motion, 
we plot $\mu_2(\alpha)-\mu_2(\pi/3)$ rather than $\mu_2(\alpha)$, with 
the zero-gap mobility $\mu_2(\pi/3)=1.739$ (see next section for the derivation).
Numerical results
are not available when
the size of the gap between two surfaces is  comparable with 
the numerical accuracy. For smaller values of $\alpha$, 
asymptotic expressions analogical to Eqs. (12)-(13) in Ref.~\cite{EW} could be derived using the same procedure. 
Notice that in the lubrication regime, the spinning speed of the beads is 
significantly larger than the angular velocity of the chain.

It remains to discuss the main issue of this paper, that is how the chains orient with time. Taking the frame of reference $(x',\;y',\;z')$, in which $\bF_0=(0,0,-1)$, and using the spherical coordinates, we parametrize the chain symmetry axis $\hat{\bf z}$ (see Fig.~\ref{skosna} for the sense of $\hat{\bf z}$) by the angles $\theta$ and $\varphi$. 
We now have $\cos \theta = -\bF_0 \cdot \hat{\bf z}$, and 
we are interested in 
the time evolution of the angle $\theta$. 
In general, it depends on the angle $\psi$ between the line of centers of spheres 1 and 3 and the unit vector $\hat{\bf e}_{\theta}=\partial \hat{\bf z}/\partial \theta$. 
 From 
Eqs.~\eqref{oO} and \eqref{rotki} we obtain,
\bee
\dot{\theta}&=& - 
(\mu _{a}\cos^2\psi+\mu
_{b}\sin^2 \psi)\sin \theta.\label{et}
\eee
For $\pi/3 < \alpha < \pi$, the calculated mobility coefficients $\mu_a$ and $\mu_b$ are positive, and therefore from Eq.~\eqref{et} it immediately follows that $\theta$ evolves towards zero, e.g. towards the chain-axis antiparallel to gravity. This orientation of the chain will be called `V-shaped'. It is the only stable orientation of the chain with 
$\pi/3 < \alpha < \pi$.

Two examples of a one-dimensional dynamics are recovered from Eq.~\eqref{et} for two 
symmetric cases with $\psi=0$ and $\psi=\pi/2$, respectively. 
For $\psi=0$, Eq.~\eqref{et} reads $\dot{\theta} = - \mu_a \sin \theta$. This dynamics 
is illustrated in Fig.~\ref{in}, with its upper C-shaped configuration, 
which corresponds to $\theta(t=0)=\pi/2$,  evolving towards the bottom one (V-shaped). 
The sense of the rotation is illustrated by the arrow. 
For $\psi=\pi/2$, Eq.~\eqref{et} reads $\dot{\theta} = - \mu_b \sin \theta$. This dynamics 
is also illustrated in Fig.~\ref{in}, now with its middle plane configuration, 
which corresponds to $\theta(t=0)=\pi/2$,  evolving towards the bottom one (V-shaped). 
The sense of the rotation is illustrated by arrows of different lengths.

To complete the analysis, we still need to compare a typical time scale of the hydrodynamic orienting to that characteristic for the settling motion. We therefore find the angles $\alpha_{a,max}$ and $\alpha_{b,max}$ which correspond to the maxima of $\mu_a$ and $\mu_b$, respectively, and then calculate the ratio of both time scales, $2\pi \mu_1/\mu_a$ and $2\pi \mu_2/\mu_b$, at $\alpha_{a,max}$ and $\alpha_{b,max}$, respectively.
For the rigid chain, both ratios are of the order of 300. For the spinning beads,  $2\pi \mu_2/\mu_b$ and $2\pi \mu_2/\mu_{\omega}$ at $\alpha_{max}$ are of the order of 150. Therefore the characteristic time scale of the hydrodynamic orienting is at least two orders of magnitude larger than that of the gravitational settling. Reorientation of the sedimenting nonsymmetric particles is significant for such systems which stay under gravity for a sufficiently long time.

\section{Stationary configurations}\label{sta}
The goal of this section is to show what are the chain configurations, which do not orient 
 under gravity. Settling speeds of such stationary configurations will be in addition compared to the translation velocities of other stationary configurations of three spheres, with the emphasis on those with the spinning particles. 

By definition, at a stationary configuration the spheres have equal translational 
velocities, 
\bee
\bU_i&=&\bU. \label{eqdef} 
\eee
Such a configuration is an equilibrium solution of the dynamics of the relative positions. 
Notice that $\bU$ is time-independent.
In our case, obviously, $i=1,2,3$.

\begin{table*}
\begin{tabular}{|l|c|c|c|c|c|c|c|}
\hline
&&&&&&&\\
&rod $||$ \bF&rod $\perp$ \bF &vertical chain&star $||$ \bF&
star $\perp$ \bF&kissing& ring \\
&&&&&&&\\
&\includegraphics[width=0.29cm]{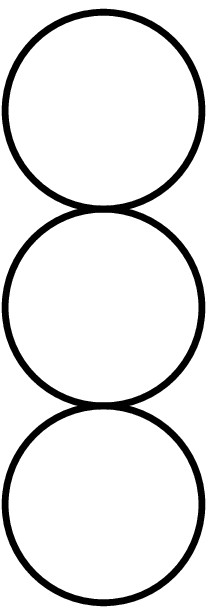}&
\includegraphics[width=0.84cm]{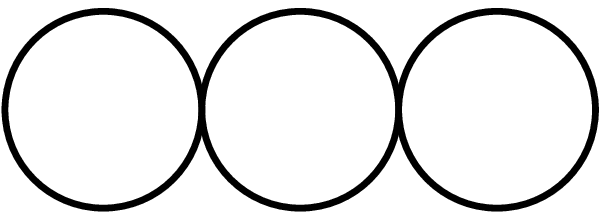}&
\includegraphics[width=2.1cm]{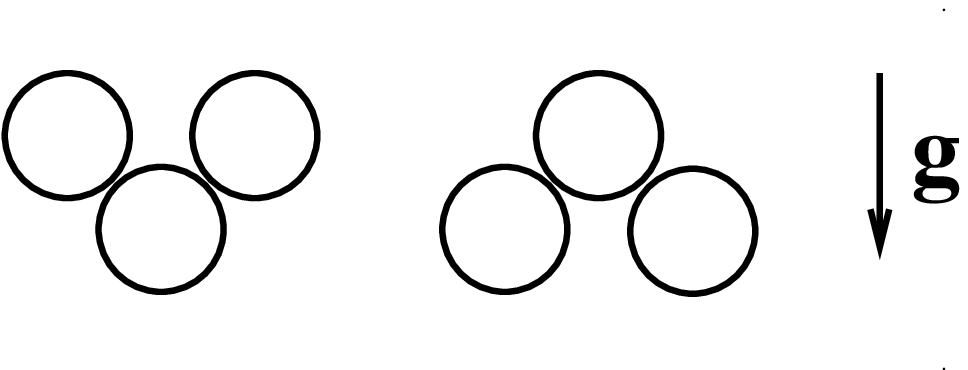}&
\includegraphics[width=0.6cm]{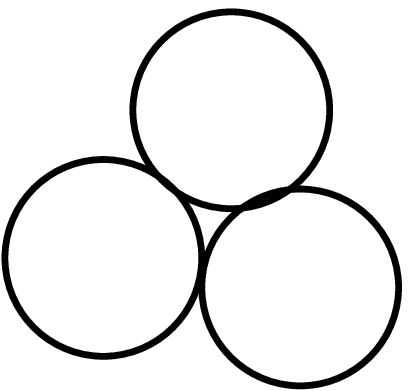}&
\includegraphics[width=0.6cm]{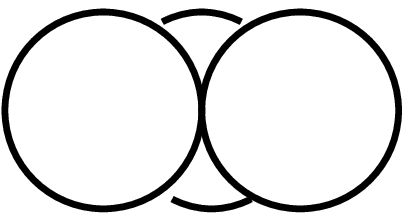}&
\includegraphics[width=1.9cm]{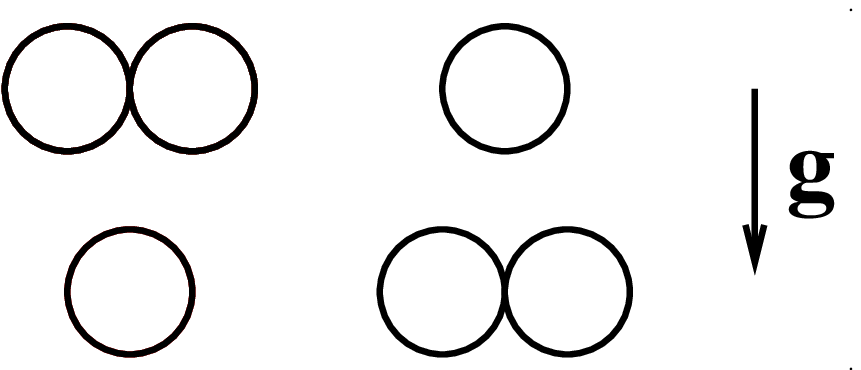}&
\includegraphics[width=1.1cm]{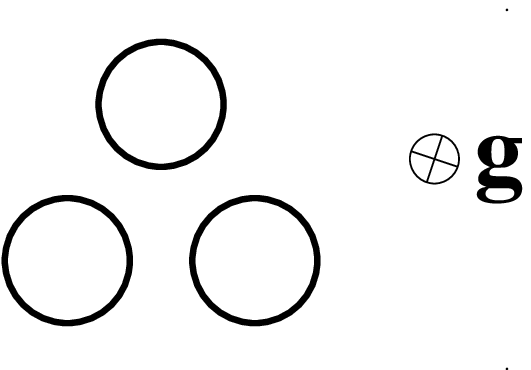}\\
&&&&&&&\\
\hline
&&&&&&&\\
spheres&1.95&1.63&1.63-1.90&1.90&1.74&1.75&1.74-1.79-1\\
\hline
point-particle&&&&&&&\\
approximation&2.00&1.61&1.61-2.04&2.04&1.75&1.81&1.75-1.74-1\\
\hline
\end{tabular}
\caption{Settling velocities $U$ of stationary configurations of three 
spheres.}\label{tabI}
\end{table*}

\subsection{Stationary chains}\label{chain}
For chains, the equilibrium condition~\eqref{eqdef} is equivalent to the relation,
\bee
\oOmega={\bf 0},
\eee
which  takes the form 
\bee
\mu_a(\alpha)\, F_x = \mu_b(\alpha)\, F_y =0,\label{wa}
\eee 
if  Eqs.~\eqref{oO} and \eqref{rotki} are applied. 
According to the numerical results plotted in Fig.~\ref{omegi}, Eq.~\eqref{wa} has the solutions,
\begin{itemize}
\item[$(i)$] an arbitrary $\alpha$, and $F_x=F_y=0$, 
\item[$(ii)$] an arbitrary $\bF$, 
and $\alpha= \pi/3,\;or \;\pi$.
\end{itemize}
The condition $(i)$ corresponds to the vertical chain equilibria, 
found in Ref.~\cite{EW} and sketched below in Fig.~\ref{no}. 
\begin{figure}[ht]
\resizebox{4.2cm}{!}{\includegraphics{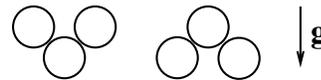}}
\caption{Vertical chain configurations.}\label{no}
\end{figure}
These are the chains with 
the symmetry axis parallel to gravity. 
The condition $(ii)$ corresponds to stars and rods, sketched in Fig.~\ref{rr}. 
\begin{figure}[ht]
\resizebox{3cm}{!}{\includegraphics{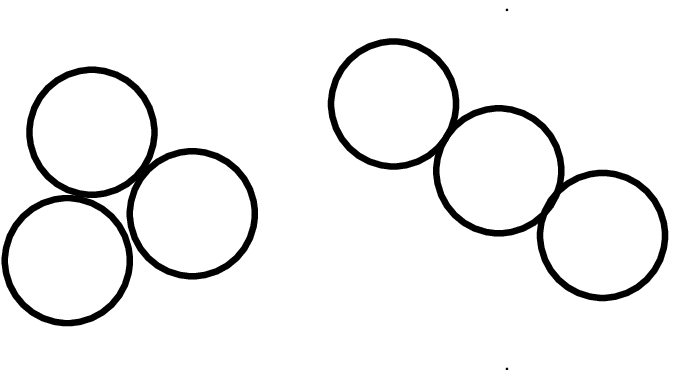}}
\caption{Star-shaped configurations with $\alpha=\pi/3$ (left) and rods with $\alpha=\pi$ (right). The direction of gravity is arbitrary.}\label{rr}
\end{figure}

Notice that from the symmetry it follows that 
the beads of the stationary chains do not spin, 
$\omega=0$. 
The settling velocities $\bU$ of the stationary chains will now be evaluated. 
The results will be also compared with the point-particle model. For touching spheres, 
such an approximation has to take into account additional constraint forces, which do not allow the points, which approximate the touching spheres, to change the interparticle distance~\cite{free}.  

Consider first the vertical chains. From symmetry with respect to reflections $x,X\rightarrow -x,-X$ and $y\rightarrow -y$ it follows that 
$\bU$ is vertical. 
Values of the settling velocities,  $U(\alpha)=\mu_3(\alpha)$, evaluated in Ref.~\cite{EW}, and replotted here in Fig.~\ref{trans}, span the range 
$\mu_3(\pi)\le U \le \mu_3(\pi/3)$, with
\bee
\mu_3(\pi)&=&1.63045819,\\
\mu_3(\pi/3)&=& 1.9022670.
\eee
Similar values follow from the point-particle approximation with constraints, 
 $ 29/18 \le U^{points} \le 229/112$.

The settling velocities $U_{\parallel}$ and $U_{\perp}$ of the rods parallel and perpendicular to gravity are now evaluated for the subsequent values of 
the multipole order $L\le 30$, and extrapolated to $L \rightarrow \infty$. Then,
\bee
U_{\parallel}=\mu_1(\pi)&=&1.946299144,\label{r1}\\
U_{\perp}=\mu_2(\pi)&=&1.63045819.\label{r2}
\eee
These values are again well approximated by the point-particle approximation with constraints, with $U_{\parallel}^{points}=2$ and 
$U_{\perp}^{points}=29/18.$ 
Both parallel and perpendicular rods settle down vertically, i.e. along gravity. 
The calculated velocities \eqref{r1}-\eqref{r2} agree with the previous experimental and numerical results~\cite{kasper,geller}.

In a similar way we calculate velocities of the stars. 
Notice that owing to the symmetry with respect to rotation by $\pi/3$, the stars located in the vertical plane settle with the same velocity, independently of their orientation. However, 
 their settling velocity $U_{\parallel}$ is larger than that of the stars oriented 
horizontally, $U_{\perp}$.  In both cases, the stars translate vertically. We evaluate,
\bee
U_{\parallel}=\mu_1(\pi/3)&=&1.90226703, \label{s1}\\
U_{\perp}=\mu_2(\pi/3)&=&1.73941260.\label{s2}
\eee
In the point-particle approximation with constraints, 
$U_{\parallel}^{points}=229/112$ and \nobreak{$U_{\perp}^{points}=7/4$}. The results \eqref{s1}-\eqref{s2} obtained for the stars improve the accuracy of the previous simulations~\cite{CH}, and agree well with the measurements~\cite{LW}.

In general, the stars and rods are inclined at a certain angle with respect to gravity. In this case, their velocities are not vertical, and the components follow from Eq.~\eqref{transki}.
A special case of such inclined stars was 
discussed in Ref.~\cite{EW}, where it was indicated by the dotted line in Fig. 15. 
In this ``slanted 
equilateral chain'' configuration, a line of centers was perpendicular to gravity.

Using the dynamics derived in the previous section, 
We conclude that the only stable 
stationary solutions of the  dynamics of chains are the V-shaped vertical chains. The
hat-shaped vertical chains are unstable. Rods are neutrally stable. So as the stars if the  three contact points are kept. The stars are unstable against perturbations which separate out a pair of the touching spheres. 

\subsection{Comparison with other stationary configurations}
The settling speeds of stationary chains will be now compared with the motion of other 
 equilibrium configurations of three spheres.
At the equilibrium, the triangle formed by the sphere centers has the following shape, size and orientation with respect to gravity.
\bi
\item ``Vertical chain'' 
(a vertical isosceles triangle with the symmetry axis along gravity; the apex sphere touches each of the base spheres).
\item ``Rod'' (a straight line of an arbitrary orientation with respect to gravity; there are  two contact points between the sphere surfaces).
\item ``Star'' (an equilateral triangle at an arbitrary orientation with respect to gravity; there are three contact points between the sphere surfaces). 
\item ``Kissing''
(an isosceles triangle with the symmetry axis along gravity and the touching 
base particles; the distance between the contact point and the center of the apex particle 
equals 1.578634 diameter, see Ref.~\cite{EW} and \footnote{Settling velocity of the kissing equilibria, $U=1.7543000$,  was
evaluated in Ref.~\cite{EW}. Here we check that it is reasonably well approximated by $U^{points}=1.814803$.}). 
\item ``Ring'' (an equilateral triangle of an arbitrary side length, in the plane perpendicular to gravity, see Refs.~\cite{hocking,caflish}).
\ei

In Table~\ref{tabI}, the stationary configurations are sketched and values of their vertical velocities are listed, together with their approximation by point-particles with constraints.
We have demonstrated that for a small number of particles, the settling velocities of their stationary configurations can be within a few percent approximated by the point particle model with constraints. Notice that all the equilibria except the ring are unstable, if arbitrary perturbations are allowed, including separation of the touching surfaces~\cite{EW}.

The ring is the only equilibrium configuration with the rotation of 
the individual spheres. It is therefore interesting to investigate 
if the spinning increases the settling velocity, as it has been observed 
for chains made of beads. This problem will be discussed in details 
in a separate  section.

\subsection{Stationary configurations with spinning}
\begin{figure}[ht]
\psfrag{U}{\LARGE $\!\!\!\!\!\!\!U$    }
\psfrag{l}{\LARGE  $\;\;\;\;\;\;\;\;\;\; \ell$}
\psfrag{ln(l-1)}{\Huge $-1/\ln(\ell-1)$}
\resizebox{8.6cm}{!}{\includegraphics{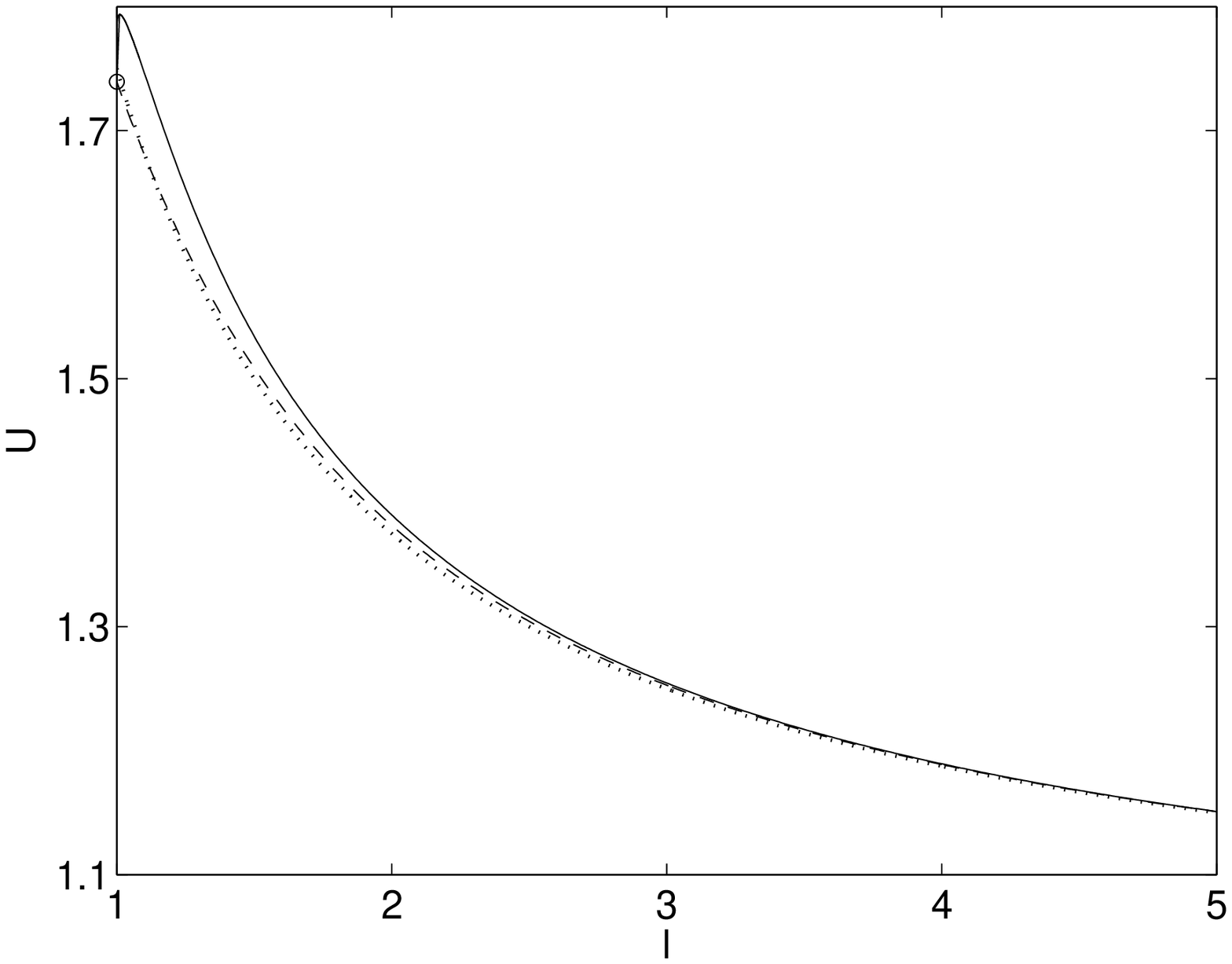}}
\begin{picture}(1,1)(25, -90)
\psfrag{U}{\Huge $\!\!\!\!\!\!\!U$    }
\resizebox{5cm}{!}{\includegraphics{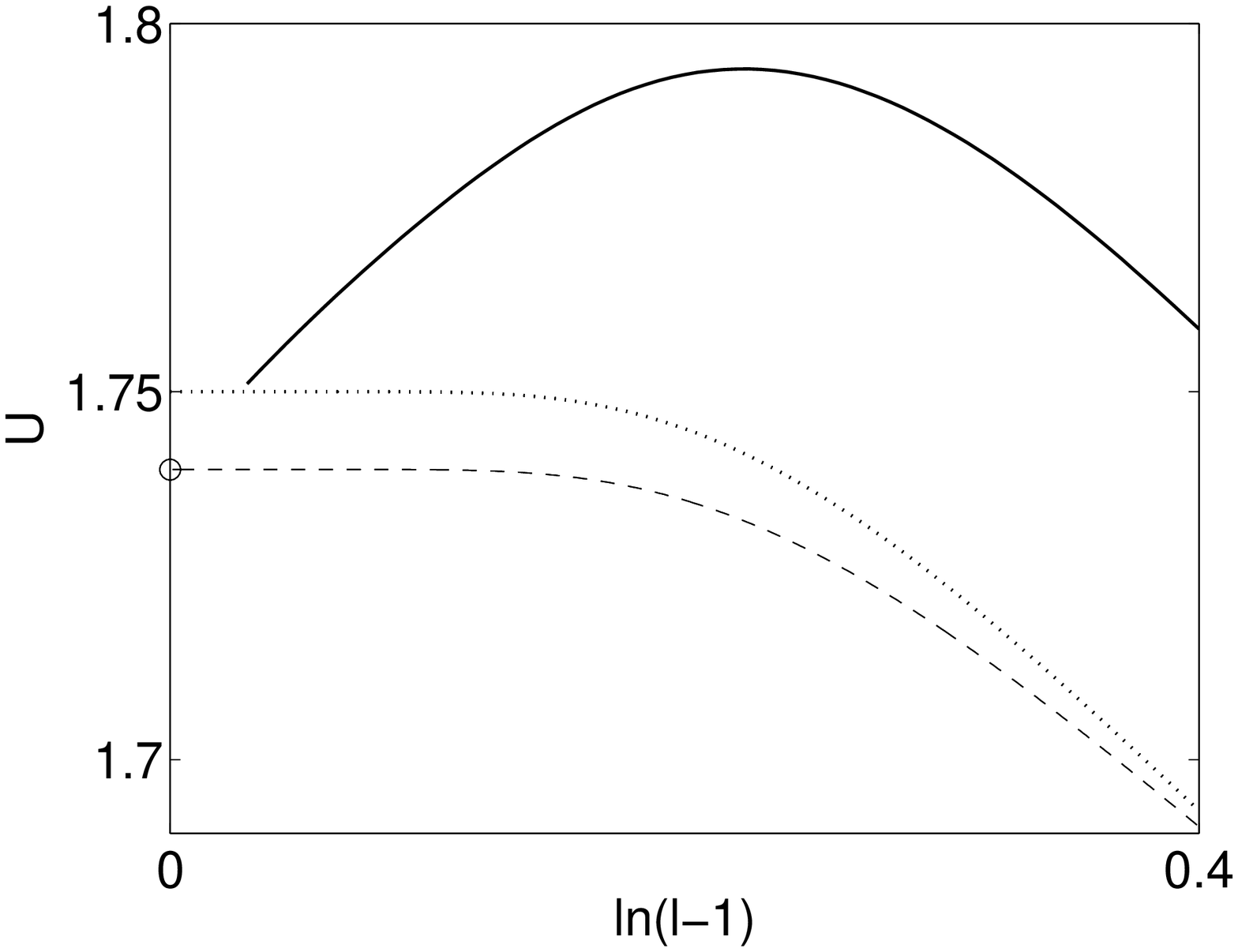}}
\end{picture}
\vspace{-0.7cm}
\caption{Settling velocity $U$ of the ring 
(equilateral horizontal triangle) versus its side length $\ell$. 
Equilibrium (solid line); rigid system with constraints (dashed line), point-particle approximation (point line) and the horizontal star ($\circ$).  Inset: $U$ as a function of 
$[-1/\ln (\ell-1)]$ for very close particles.}\label{U}
\end{figure}

We now focus on the stationary configurations called rings. The sphere centers 
form a horizontal equilateral triangle with an arbitrary length $\ell\ge 1$ of its side.
Settling velocity $\bU$ is of course vertical. 
Its value $U=|\bU|$ is evaluated and plotted in Fig.~\ref{U} as a function of $\ell$. 
In general, the particles are separated from each other; they touch only 
in the limiting case of $\ell=1$, when the ring becomes the horizontal star, with $U=1.73941260$. 

It is interesting to observe that the ring's settling velocity has a maximum 
for a very small gap 
between the sphere surfaces. The maximum is well-visible at the inset of Fig.~\ref{U}. Bracketing 
this maximum by the standard golden 
section search~\cite{NR}, we evaluate the corresponding values 
of $\ell_{MU}$ and $U(\ell_{MU})$. The subsequent 
multipole orders $1\le L \le 28$ are used, and the results are next extrapolated to  $L \rightarrow \infty$, as in Ref.~\cite{EW}. We obtain, 
$\ell_{MU} = 1.01128$ and 
$U(\ell_{MU}) = 1.79394$. 
For larger values of $\ell$, the settling velocity decreases to zero 
with $\ell \rightarrow \infty$. In Tab.~\ref{tabI}, the ring velocities at $\ell=0,\;\ell_{MU}$ and $\infty$ have been indicated. 

To check if the existence of the maximum is related to the spinning, 
we also consider a rigid system of three spheres at the same configurations, 
but with the spinning eliminated owing to constraint external torques. 
Velocities of such configurations are also 
plotted in Fig.~\ref{U}. They are systematically smaller than the velocities 
of the ring. Indeed, this example also indicates that spinning speeds up 
the rate of settling by a small amount. 

The point-particle approximation, 
$U=1+3/(4\ell)$, is also depicted in Fig.~\ref{U}.
Notice that for rings, no constraint forces nor torques 
are applied, since the rings are also stationary solutions 
of the point-particle dynamics. It is clear that the point particle 
approximation is much closer to the rigid dynamics than to the spinning system.

\begin{figure}[ht]
\psfrag{l}{\LARGE  $\;\;\;\;\;\;\;\;\;\; \ell$}
\psfrag{OO}{\LARGE $\;\;   \Omega_o$}
\resizebox{8.6cm}{!}{\includegraphics{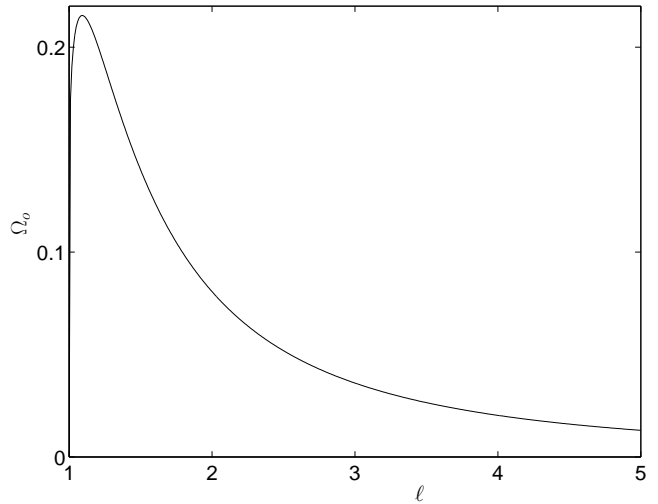}}
\caption{Spinning velocity $\Omega_o$ versus the size $\ell$ of the ring.}\label{O}
\end{figure}

Finally, we evaluate the spinning velocities of the spheres.
Here $\oOmega_1\!=\!\Omega_o \,\hat{\br}_{32}$, 
$\oOmega_2\!=\!\Omega_o \,\hat{\br}_{13}$, 
and $\oOmega_3\!=\!\Omega_o \,\hat{\br}_{21}$, 
with $\Omega_o$, 
plotted in Fig.~\ref{O} as a function of the side 
length $\ell$. The 
maximum of $\Omega_o$ is reached at the small distance between the sphere centers,
$\ell_{M\Omega}=1.0923791$, but not as small as $\ell_{MU}$. At the maximum, the 
spinning velocity
$\Omega_o(\ell_{M\Omega}) = 0.215537174$, is as much as 12\% of 
the settling speed,
$U(\ell_{M\Omega})= 1.7519797$.   
At the maximum of $U$, 
the spinning is slightly smaller, with $\Omega_o(\ell_{MU})=0.17704.$ 
These two maxima are shifted with respect to each other, because  $U$ 
becomes larger  not only by an increase of $\Omega_o$, but also by a decrease 
of the distance between the spheres. 

Notice that for the ring configuration of the separated spheres, 
the maximal spinning velocity is  two times larger than for the horizontal chain of the touching beads. As a consequence, the increase of the settling velocity is also twice as large. 

\section{Conclusions}\label{con}
The goal of this paper has been to construct and study a simple model of a chain-like asymmetric microobject of a fixed shape, 
settling under gravity in a viscous fluid. The motion of such a system has been evaluated from the multipole expansion of the Stokes equations. 
The main results are the following.

It has been found that asymmetric microobjects orient hydrodynamically while settling
under gravity. This effect is not observed for axially symmetric objects such as rods, 
neither for regular shapes such as equilateral triangles, here called stars.
However, chain-like conglomerates made of two identical straight arms in general orient
towards a vertical ``head down'' equilibrium configuration; that is, towards a V-shape. This process is relatively slow in comparison to the cluster sedimentation, but definitely not negligible. We have checked that the hydrodynamic orienting is observed for chains made of a central particle with the attached two identical straight arms made of a certain number of spinning or rigid beads, for example ten spheres at each arm. The central particle may be the same size as the other ones, or larger, e.g. 10 times larger.

It has been shown that freely rotating particles in chain-like conglomerates can spin even if their surfaces touch each other. The spinning can be even faster than the chain rotation. 
The spinning particles speed up the conglomerate settling by a few percent, and they significantly  
enhance its tendency to orient vertically, 
in comparison to the rigid body. Spinning also speeds up the settling of stationary configurations of spheres separated from each other. 

Naturally, the hydrodynamic orienting found in this work is important for efficient swimming of microorganisms which are more dense than the fluid. The results are also relevant for  
suspensions of chain-like conglomerates settling under gravity. 
On the long time scale, while reorienting takes place, the suspension structure and settling speed may change, leading to ordering of the sediment and possible applications to segregation and filtration techniques.

\appendix
\section{How to evaluate mobility of a conglomerate?}\label{collective}
For the system of $N$ particles separated from each other,
the $6\times 6$ mobility matrices $\m_{ij}$ form a $6N\times 6N$ tensor, which is evaluated as 
the inverse 
of the $6N\times 6N$ friction tensor, made of 
$6\times 6$ friction matrices $\z_{ij}$. The latter  
relate the external forces and 
torques, $\bF_i$ and $\bT_i$,  exerted on a particle $i=1,...,N$, to the translational and angular velocities, $\bU_j$ and $\oOmega_j$, of a particle $j=1,...,N$,
\begin{equation}
\left( 
\begin{array}{c}
\bF_{i} \\ 
\bT_{i}%
\end{array}%
\right) \mathbf{=}\sum_{j=1}^{N}
\z_{ij}
\cdot \left( 
\begin{array}{c}
\bU_{j} \\ 
\oOmega_{j}%
\end{array}%
\right) .  \label{002}
\end{equation}%

In a conglomerate, the particles touch each other, and some of the $\z_{ij}$ components become infinite. Instead of using Eq.~\eqref{002},  it is therefore necessary to eliminate the forbidden degrees of freedom (relative motions). It is done by constructing 
the conglomerate friction as the sum of the relevant combinations of $\z_{ij}$ only,
\begin{equation}
\z = \sum_{i=1}^{N}\sum_{j=1}^{N}\mathbf{P}_{i}^{T}\cdot \z_{ij}\cdot \mathbf{P}_{j}.  \label{011}
\end{equation}%
The $6\times 6$ friction matrix $\z$ is finite at the contact, because the combinations 
of $\z_{ij}$ in Eq.~\eqref{011} correspond to motions, which are free from 
the lubrication
singularities. By inverting $\z$, we obtain the conglomerate mobility,
\bee
\m&=&\z^{-1}.\label{mz}
\eee

In the following, we construct the operators $\mathbf{P}_{i}$ for two different types of the conglomerates. 
First, a rigid system is considered, for which all relative 
motions are excluded. Then, a conglomerate with spinning particles is analyzed.

\subsection{Hydrodynamics of a rigid system}
A rigid motion of a conglomerate made of $N$ spheres 
is characterized
by  the translational velocity $\bU$ of an arbitrary chosen center
of reference $\mathbf{R}$ and the conglomerate rotational velocity $\mathbf{%
\Omega }$. Then, the translational and rotational velocities of the individual sphere centers,  $i=1,\ldots ,N$, are
given~as %
\begin{eqnarray}
\bU_{i} &=&\bU+\oOmega \times (\mathbf{r}_{i}-\mathbf{R}),
\label{003} \\
\oOmega_{i} &=&\oOmega,
\end{eqnarray}
or equivalently,
\begin{equation}
\left( 
\begin{array}{c}
\bU_{i} \\ 
\oOmega_{i}%
\end{array}%
\right) =\mathbf{P}_{i}\cdot \left( 
\begin{array}{c}
\bU \\ 
\oOmega%
\end{array}%
\right) \mathbf{,}~~~i=1,\ldots ,N,  \label{004}
\end{equation}%
where the $6\times 6$ matrices $\mathbf{P}_{i}$ are given by the relation,%
\bee
\mathbf{P}_{i}&=&\left( 
\begin{array}{cc}
{\bf I} & \mathbf{P}_{i}^{tr} \\ 
0 & {\bf I}%
\end{array}%
\right) ,  \label{005}\\
(\mathbf{P}_{i}^{tr})_{\alpha \beta } &=&\varepsilon _{\alpha \beta \gamma
}(r_{i\gamma }-R_{\gamma }).  \label{ptr}
\eee%
The total force $\bF$\ and torque $\bT$ with respect to the center $\mathbf{R}$, exerted externally on the conglomerate, have the form, %
\begin{eqnarray}
\bF &=&\sum_{i=1}^{N}\bF_{i},  \label{007} \\
\bT &=&\sum_{i=1}^{N}[\bT_{i}+(\mathbf{r}_{i}-\mathbf{R})
\times \bF_{i}],\label{077}
\end{eqnarray}%
or equivalently,
\begin{equation}
\left( 
\begin{array}{c}
\bF \\ 
\bT%
\end{array}%
\right) \mathbf{=}\sum_{i=1}^{N}\mathbf{P}_{i}^{T}\cdot \left( 
\begin{array}{c}
\bF_{i} \\ 
\bT_{i}%
\end{array}%
\right),  \label{008}
\end{equation}%
where $^T$ stands for the matrix transposition. %
By inserting Eqs.~(\ref{002}) and (\ref{004}) into Eq. (\ref{008}), 
we relate the total external force and torque on the conglomerate to its 
translational and rotational velocities,%
\begin{equation}
\left( 
\begin{array}{c}
\bF \\ 
\bT%
\end{array}%
\right) \mathbf{=}
\begin{array}{cc}
\z
\end{array}%
\cdot 
\left( 
\begin{array}{c}
\bU \\ 
\oOmega%
\end{array}%
\right),  \label{010}
\end{equation}%
with the conglomerate friction $\z$ given by Eq.~\eqref{011}. 
Writting the conglomerate velocities explicitly it terms of the corresponding components of the mobility $\m=\z^{-1}$, 
\begin{equation}
\left( 
\begin{array}{c}
\bU \\ 
\oOmega%
\end{array}%
\right) \mathbf{=}\left( 
\begin{array}{cc}
\m^{tt} & \m^{tr} \\ 
\m^{rt} & \m^{rr}%
\end{array}%
\right) \cdot \left( 
\begin{array}{c}
\bF \\ 
\bT%
\end{array}%
\right),  \label{012}
\end{equation}
and choosing the center of mass as the reference center, we obtain $\bT={\bf 0}$, and we recover the relations \eqref{uU}-\eqref{oO}, 
with the rigid-chain mobility matrices $\m^{tt}$ and $\m^{rt}$. 

\subsection{Hydrodynamics of a chain made of beads}
Let us now consider a chain of three beads. The sphere 2 touches the other spheres, but the 
 spheres 1 and 3 are separated from each other and therefore are able to spin along $\hat{\br}_{12}$\ and\ $\hat{\br}_{32}$, respectively. The motion 
of the chain is characterized by 
the translational velocity $\bU$ of an arbitrary center of reference $%
\mathbf{R}$, the chain rotational velocity $\oOmega$, and also
by the two spinning velocities $\omega _{1}$ and 
$\omega _{3}$. 
Then, the 
translational and rotational velocities of individual sphere centers are given by the relation,%
\begin{eqnarray}
\bU_{i} &=&\bU+\oOmega\times (\mathbf{r}_{i}-\mathbf{R%
}),~~~~i=1,2,3, \hspace{0.5cm} \label{013} \\
\oOmega_{i} &=&\mathbf{\Omega +}\omega _{i}\hat{\br}%
_{i2},~~~~\hspace{1.4cm} i=1,3, \\
\oOmega_{2} &=&\oOmega,
\end{eqnarray}%
or equivalently,
\begin{eqnarray}
\left( 
\begin{array}{c}
\bU_{i} \\ 
\oOmega_{i}%
\end{array}%
\right)  &=&\mathbf{P}_{i}\cdot \left( 
\begin{array}{c}
\bU \\ 
\oOmega \\ 
\omega _{1} \\ 
\omega _{3}%
\end{array}%
\right),  \hspace{0.6cm} i=1,2,3,\label{014} 
\end{eqnarray}%
with the $6\times 8$ matrices $\mathbf{P}_{i}$ defined with the use of the same Eq.~\eqref{ptr} for 
$\mathbf{P}_{i}^{tr}$, but now 
 differently than in 
Eq.~(\ref{004}), 
\bee
\mathbf{P}_{1}&=&\left( 
\begin{array}{cccc}
{\bf I} & \mathbf{P}_{1}^{tr} & 0 & 0 \\ 
0 & {\bf I} & \hat{\br}_{12} & 0%
\end{array}%
\right),  \label{015} \\ \nonumber 
\eee

\newpage
\bee
\mathbf{P}_{2}&=&\left( 
\begin{array}{cccc}
{\bf I} & \mathbf{P}_{2}^{tr} & 0 & 0 \\ 
0 & {\bf I} & 0 & 0%
\end{array}%
\right) \mathbf{,}  \label{116}\\
\mathbf{P}_{3}&=&\left( 
\begin{array}{cccc}
{\bf I} & \mathbf{P}_{3}^{tr} & 0 & 0 \\ 
0 & {\bf I} & 0 & \hat{\br}_{13}%
\end{array}%
\right). \label{117}
\eee%
Then, Eq.~\eqref{011} is used 
to evaluate  the $8\times 8$ chain 
friction matrix $\z$. It
relates the 
the chain velocities $\bU$, $\oOmega$, $\omega_1$ and $\omega_2$, 
to the total external forces and torques \eqref{007}-\eqref{077}, 
and the torque components $t_{1}\!=\!\bT_1 \cdot \hat{\br}_{12}$ and 
$t_{3}\!=\!\bT_3 \cdot \hat{\br}_{32}$,
\begin{equation}
\left( 
\begin{array}{c}
\bF \\ 
\bT \\ 
t_{1} \\ 
t_{3}%
\end{array}%
\right) =\z \cdot \left( 
\begin{array}{c}
\bU \\ 
\oOmega \\ 
\omega _{1} \\ 
\omega _{3}%
\end{array}%
\right)   \label{016}
\end{equation}%
Evaluating the chain mobility\ $\m=\z^{-1}$, we 
obtain,%
\begin{equation}
\left( 
\begin{array}{c}
\bU \\ 
\oOmega \\ 
\omega _{1} \\ 
\omega _{3}%
\end{array}%
\right) = \m \cdot \left( 
\begin{array}{c}
\bF \\ 
\bT \\ 
t_{1} \\ 
t_{3}%
\end{array}%
\right).  \label{017}
\end{equation}
In our system, the spheres are identical, $\mathbf{R}$ is the center-of-mass position, $\bT={\bf 0}$ and $t_{1}=t_{3}=0$.
Therefore, $\omega_1=-\omega_3=\omega$ and 
we obtain the relations \eqref{uU}-\eqref{o}.

\end{document}